\documentclass[12pt]{article}

\usepackage{scicite}
\usepackage{graphicx}

\usepackage{times}

\newcommand{\tot}[1]{\int {#1}\,{\rm d}x{\rm d}y}
\newcommand{\avr}[1]{\langle{#1}\rangle}

\newenvironment{sciabstract}{%
\begin{quote} \bf}
{\end{quote}}

\newcounter{lastnote}
\newenvironment{scilastnote}{%
\setcounter{lastnote}{\value{enumiv}}%
\addtocounter{lastnote}{+1}%
\begin{list}%
{\arabic{lastnote}.}
{\setlength{\leftmargin}{.22in}}
{\setlength{\labelsep}{.5em}}}
{\end{list}}

\title{Penumbral structure and outflows in simulated sunspots} 

\author
{M. Rempel$^{1\ast}$, M. Sch{\"u}ssler$^{2}$, 
 R.H. Cameron$^{2}$ \& M. Kn{\"o}lker$^{1}$\\
\\
\normalsize{$^{1}$High Altitude Observatory,
    NCAR, P.O. Box 3000, Boulder, Colorado 80307, USA}\\
\normalsize{$^{2}$Max-Planck-Institut f\"ur Sonnensystemforschung,}\\
\normalsize{Max-Planck-Str. 2, 37191 Katlenburg-Lindau, Germany}\\
\\
\normalsize{$^\ast$To whom correspondence should be addressed; 
                   E-mail:  rempel@hao.ucar.edu.}
}

\date{}

\begin{document} 

\maketitle 

\begin{sciabstract}
Sunspots are concentrations of magnetic field on the visible solar surface
that strongly affect the convective energy transport in their interior and
surroundings. The filamentary outer 
regions (penumbrae) of sunspots show systematic radial outward flows along 
channels of nearly horizontal magnetic field. These flows were discovered
100 years ago and are present in all fully developed sunspots.
Using a comprehensive numerical simulation of a sunspot pair, we show that 
penumbral structures with such outflows form when the average magnetic field 
inclination to the vertical exceeds about 45 degrees. The systematic outflows 
are a component of the convective flows that provide the upward energy 
transport and result from anisotropy introduced by the presence of the 
inclined magnetic field. 

\end{sciabstract}

Sunspots are dark patches on the visible solar surface that harbor
strong magnetic fields up to 4000 G \cite{Solanki:2003,
Thomas:Weiss:2004}.  Their central region, the umbra, is the darkest
part with a brightness of about 20\% of the ambient value and a largely
vertically oriented magnetic field; the brighter, filamentary penumbra
shows a more inclined field and a nearly horizontal plasma outflow of
several km$\cdot$s$^{-1}$, the Evershed flow, named after its discoverer
\cite{Evershed:1909}. While a number of simplified (and partly
conflicting) models have been suggested to explain the structure and
outflows of penumbrae \cite{Thomas:Weiss:2008}, a comprehensive 
theoretical understanding of the basic mechanisms does not exist .

Here we present ab-initio numerical simulations of complete sunspots
embedded in a realistic solar convection zone and atmosphere, including
all relevant physical processes: compressible magnetohydrodynamics,
partial ionization, and radiative energy transport.  Previous attempts
to simulate penumbral structure in small slab-like sections of sunspots
\cite{Heinemann:etal:2007, Rempel:etal:2009} resulted in rather
narrow penumbral regions. The generally used periodic boundary
conditions at the sides of the computational box tend to suppress the
extended horizontal field structures associated with sunspot penumbrae. Hence
we have carried out a simulation of a pair of big sunspots (diameter ~35
Mm) of opposite magnetic polarity, thereby facilitating the development of
strongly inclined field between the spots.  Our numerical box had a
horizontal extension of $98\,$Mm$\times 49\,$Mm and a depth of
$6.1\,$Mm. The spatial grid resolution was $32\,$km in the horizontal
directions and $16\,$km in the vertical. The sunspots evolved
for $3.6$ hours during the simulation, which was sufficient to study the
penumbral structure and dynamics; processes that evolve on longer time 
scales such as moat flows were not fully developed in this simulation. 
However, the surface evolution of magnetic field shows clear
indications of bipolar magnetic features transported away from
the spots beyond the penumbra boundary. This is reminiscent to observations 
of so-called `moving magnetic features' \cite{Kubo:etal:2007} 
(SOM, movie1.mpg: 'Magnetogram' 
movie displaying the temporal evolution of $B_z$ on the visible solar surface, 
the gray scale is ranging from $-3.5$~kG to $+3.5$~kG). More detailed 
information about the physical model, the numerical code, and the simulation 
setup is provided in the SOM.

The simulated penumbrae show the largest extension between the sunspots of
opposite polarity (Fig.~1A). The periodic horizontal boundary conditions 
provide three different distances: $46$~Mm (middle of box) and $52$~Mm
in the $x$-direction between opposite polarity spots, and $49$~Mm in the 
$y$-direction between same polarity spots. The spots show a dark umbra 
with some brighter umbral dots, preferentially in the weaker spot on the 
left. A deep magnetic structure underlies the visible penumbra, 
particularly so between the sunspots (Fig.~1B). A movie covering $1$ hour
of temporal evolution of the properties displayed in Fig.~1 (SOM, movie2.mpg) 
shows the inward progression of filaments in the inner penumbra.

The umbral regions have a brightness of $0.15$ to $0.2\,I_0$, where $I_0$ is 
the average quiet-Sun value, a Wilson depression of the visible surface 
by 550--600~km, and vertical field strength ($B_z$) up to 4~kG (Fig.~2).
The quantities described here are averaged in space and time as described
in the caption of Fig.~2. 
The penumbrae have much weaker $B_z$, horizontal fields ($B_x$)
with peak values around 2~kG at the inner penumbral boundaries, and an
average brightness of about $0.7\,I_0$. The penumbral region
exhibit systematic outflows with average horizontal velocities ($v_x$)
of up to 6~km$\cdot$s$^{-1}$. The onset of these flows is closely
related to the magnetic field inclination: where the average
inclination with respect to the vertical exceeds 45 deg, there are
systematic average outflows.  With
growing distance from the umbra, the outflow velocity
increases and the field becomes more inclined and is
nearly horizontal in the outer penumbra.  These properties are
consistent with observational results \cite{Keppens:Martinez:1996,
Solanki:2003}.

The simulated penumbra shows strong structuring in terms of elongated
narrow filaments (Fig.~3). In the inner part the magnetic field shows 
strong variations of the inclination between 40 deg and nearly 
90 deg on scales of less than 200 km. Further out regions with strongly 
inclined field dominate. The velocity structure is analogous: radial 
outflows are concentrated in highly inclined filaments 
and become stronger and azimuthally more extended in the outer penumbra,
where the field is almost uniformly horizontal. Vertical (up- and
downward) flows occur in narrow filaments throughout the whole
penumbral region.

Analyzing the penumbral structure in vertical cuts we find that 
the outflows reach their peak velocities (exceeding 10~km$\cdot$s$^{-1}$)
near the visible surface (Fig.~4). This reflects the
strong height gradient of pressure and density in these layers: rising
hot plasma turns over and the resulting horizontal flow is guided
outward from the spot by the strong and inclined magnetic field. 
While the vertical field in the sunspot umbra only permits convection
in the form of narrow columnar structures \cite{Schuessler:Voegler:2006},
the inclined field in the penumbra favours sheet-like upflows, which
are radially extended and narrow in the azimuthal direction 
\cite{Rempel:etal:2009}. Together with the preferred weakening of the vertical
field component due to flux expulsion by the expanding rising plasma,
this explains the azimuthal structuring and large azimuthal variations of 
the field inclination in the inner penumbra. The influence of horizontal flows 
on the field structure depends on the location in the penumbra. In the
inner penumbra they remain rather weak and have therefore only a limited back 
reaction on the field structure. In the outer penumbra they become 
sufficiently strong to bend over field lines leading to more extended patches 
of horizontal field and flows. In
addition to the strong localized outflows near the visible surface,
there is a large-scale flow cell with plasma rising and diverging around
the spot, as is evident by the general reddish color in the
representation of the horizontal velocity in Fig.~4.  
Systematic inflows (comprising very little mass flux) are apparent in the
uppermost layers of the simulation box. Because these regions are strongly
affected by the upper (closed) boundary, it is not clear whether the
inflows could possibly be related to the observed inverse Evershed
effect in the chromosphere \cite{Dialetis:etal:1985}.

The central penumbral region between the spots has a mean bolometric
brightness of $I_{\rm p}=0.68 I_0$ (averaged over $y=\pm 3.2\,$Mm from
the midplane of the computational box and between the central vertical dotted 
lines indicated in Fig. 2). The mean brightness of the upflow areas is 
$1.1 I_{\rm p}$, while that of downflows areas is $0.92 I_{\rm p}$. The 
corresponding values for undisturbed granulation are $1.11 I_0$ and $0.88 I_0$,
respectively, implying they have similar properties. However, the penumbral 
region shows an rms bolometric brightness contrast of
25.2\%, which is substantially larger than the corresponding granulation
value of 17.3\%. Observations also imply a positive correlation between 
brightness and vertical flow direction \cite{Almeida:etal:2007}. This 
constitutes evidence for a convective flow pattern that 
transports the energy flux emitted in the penumbra.
Other studies show a correlation 
between intensity and line-of-sight velocities \cite{Bellot-Rubio:etal:2006},
which for sunspots observed outside the center of the solar disk is dominated 
by the horizontal Evershed flow. This is consistent
with our findings, because in the penumbra the horizontal flow velocity is 
correlated with the vertical flow direction.

Our detailed analysis (SOM) shows
that the spatial scales of the flows providing the major part of the
convective energy transport are similar for both undisturbed granulation
and penumbra. The primary difference is that there is no
preferred horizontal direction for granulation, while the
energy-transporting flows in the penumbra are distinctly asymmetric:
convective structures are elongated in the radial direction of the
sunspot. These properties were already indicated in earlier simulations 
\cite{Heinemann:etal:2007, Rempel:etal:2009} and suggested as an explanation
for the Evershed outflow in \cite{Scharmer:etal:2008}. The simulation shown 
here confirms this suggestion and demonstrates the convective nature of a 
fully developed penumbra. 

The horizontal asymmetry of the convective flows is also manifest in the
correlation of $0.42$ between the corresponding flow component ($v_x$) and 
the brightness. We find that the rms of the outflowing velocity component
$(v_x)$ in the penumbra is much larger than the transverse component $(v_y$)
(perpendicular to the filament direction), showing an asymmetry similar to 
that found by the scale analysis. The total rms
velocity profile as a function of depth is very similar to its
counterpart for undisturbed granulation, apart from a slightly higher
peak value, confirming the physical similarity of convection in
granulation and penumbra. 

The mass flux and energy flux show similar properties with respect to 
the length scales and asymmetry (SOM), indicating that most 
of the outflowing material emerges, turns over and
descends within the penumbra. In the
deeper layers, there is some contribution (of the order of 10--20\%) to
both energy and mass flux by the large-scale flow cell surrounding the
sunspots.

The analysis of our simulations clearly indicates that
granulation and penumbral flows are similar with regard to energy
transport; the asymmetry between the horizontal directions and the
reduced overall energy flux reflect the constraints imposed on the
convective motions by the presence of a strong and inclined magnetic
field. The development of systematic outflows is a direct consequence
of the anisotropy and the similarities between granulation and penumbral 
flows strongly suggest that driving the Evershed flow does not require
physical processes that go beyond the combination of convection and
anisotropy introduced by the magnetic field. Weaker laterally overturning
flows perpendicular to the main filament direction explain the
apparent twisting motions observed in some filaments
\cite{Ichimoto:etal:2007, Zakharov:etal:2008} and lead to a weakening of 
the magnetic field in the flow channels through flux expulsion 
\cite{Rempel:etal:2009}. 

Although our simulations of large sunspots is realistic in terms of 
relevant physics, it does not faithfully reproduce all aspects of the 
morphology of observed penumbral filaments. The penumbral regions are
considerably more extended than in previous local simulations, but they
are still somewhat subdued, probably owing to the proximity of the
periodic boundaries. The filaments in the
inner penumbrae appear to be too fragmented and short, dark lanes along
bright filaments \cite{Scharmer:etal:2002} form only occasionally, likely
a consequence of the still limited spatial resolution of the
simulation. Finally, the initial condition
of the magnetic field underlying the sunspot is quite arbitrary,
owing to our ignorance of the subsurface structure of
sunspots. Notwithstanding these limitations, the present simulations are
consistent with observations of global sunspot
properties, penumbral structure, and systematic radial outflows.
These and earlier simulations \cite{Schuessler:Voegler:2006,
Heinemann:etal:2007, Rempel:etal:2009} suggest a unified physical
explanation for umbral dots as well as inner and outer penumbrae in
terms of magneto-convection in a magnetic field with varying
inclination. Furthermore a consistent physical picture of all 
observational characteristics of sunspots and their surroundings is
now emerging.

\begin{scilastnote}
\item High-performance computing resources were
provided by NCAR's Computational and Information Systems Laboratory. The 
National Center for Atmospheric Research is sponsored by the National
Science Foundation.
\end{scilastnote}

\begin{figure}[ht]
      \begin{center}
\resizebox{\hsize}{!}{\includegraphics{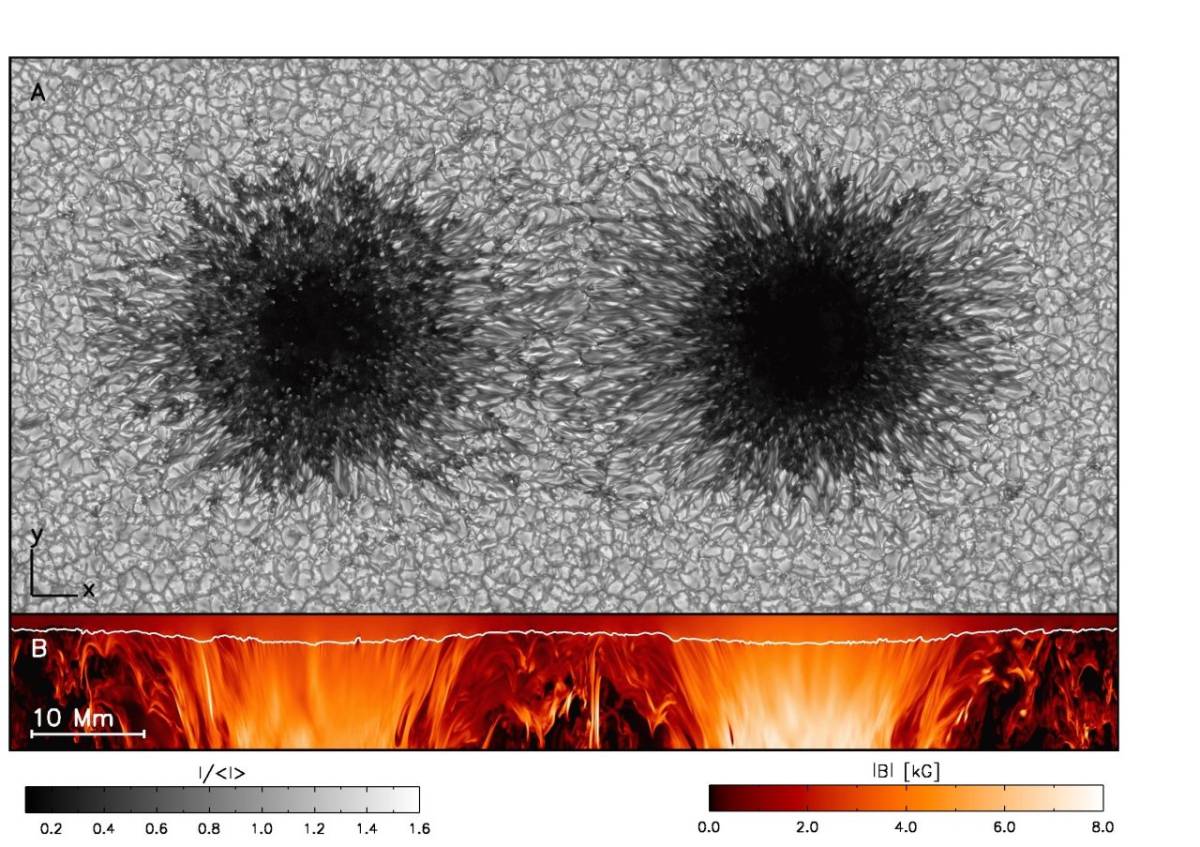}}
      \end{center}
      \caption{
        Snapshot from the simulation. ({\bf A}) Surface
        brightness map of the sunspot pair and the surrounding
        convective pattern (granulation). ({\bf B}) Color
        representation of the field strength (saturated at 8 kG)
        in a vertical cut through the midplane of the simulation
        box at $y=25\,$Mm. The vertical
        direction is stretched by a factor of 2. The white line
        indicates the height level of the visible surface
        (optical depth unity).
      }
\end{figure}

\clearpage

\begin{figure}[ht]
      \begin{center}
      \vglue -3cm
      \resizebox{\hsize}{!}{\includegraphics{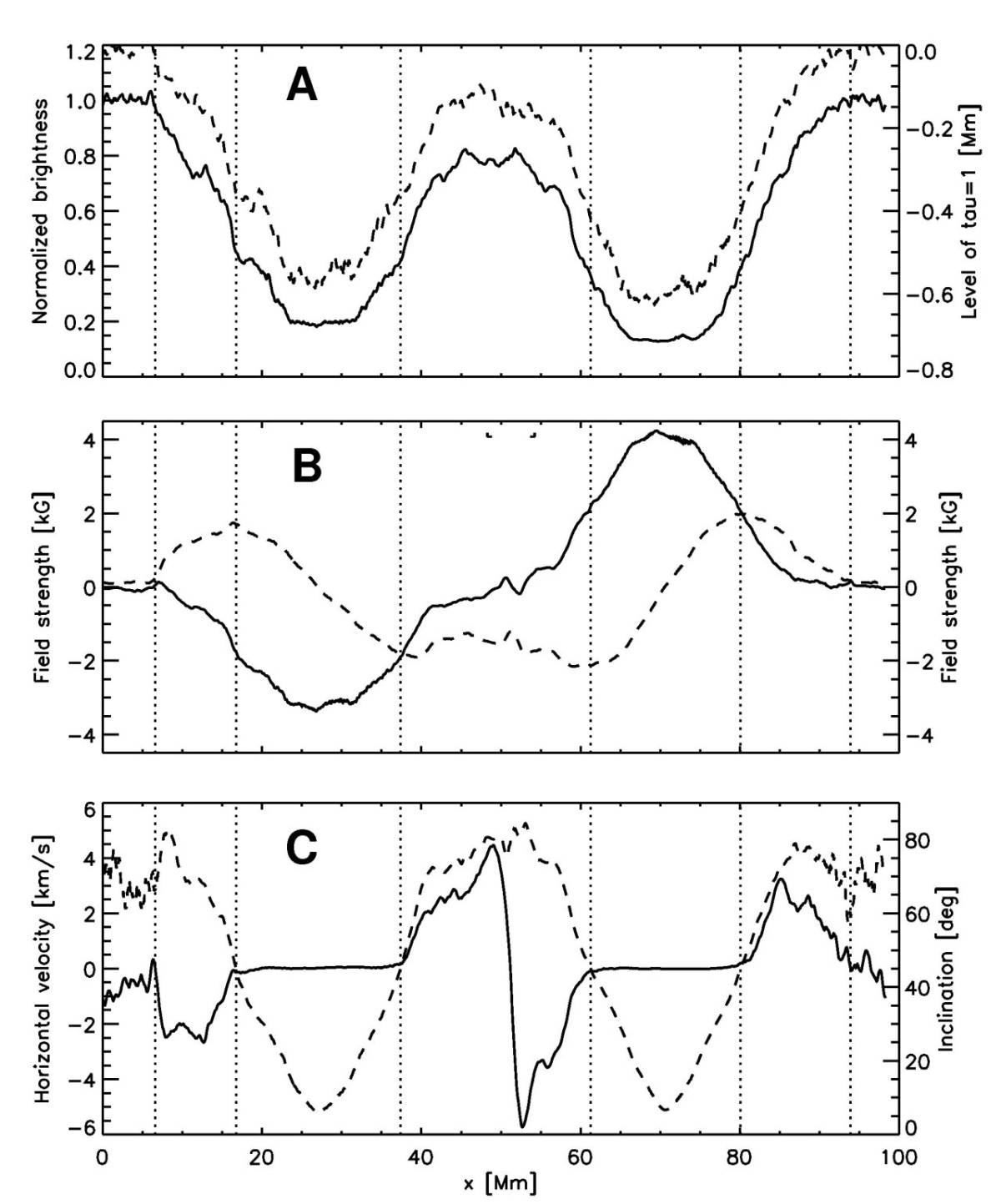}}
      \end{center}
      \caption{
        Horizontal profiles of various quantities at
        the visible surface, averaged over $\pm 3.2\,$Mm from the
        midplane in $y$ and time-averaged over one hour. ({\bf A})
        Brightness (normalized to the average over non-magnetic
        regions, solid) and depression of the visible surface
        with respect to non-magnetic regions (Wilson depression,
        dashed). ({\bf  B}) Strength of the vertical ($B_z$, solid) and
        horizontal ($B_x$, dashed) magnetic field components. ({\bf C})
        Horizontal velocity ($v_x$, solid) and magnetic field
        inclination with respect to the vertical
        (dashed). Vertical dotted lines indicate the regions with
        systematic penumbral outflows.
        }
\end{figure}      

\clearpage

\begin{figure}[ht]
        \vglue -2cm
        \resizebox{\hsize}{!}{\includegraphics{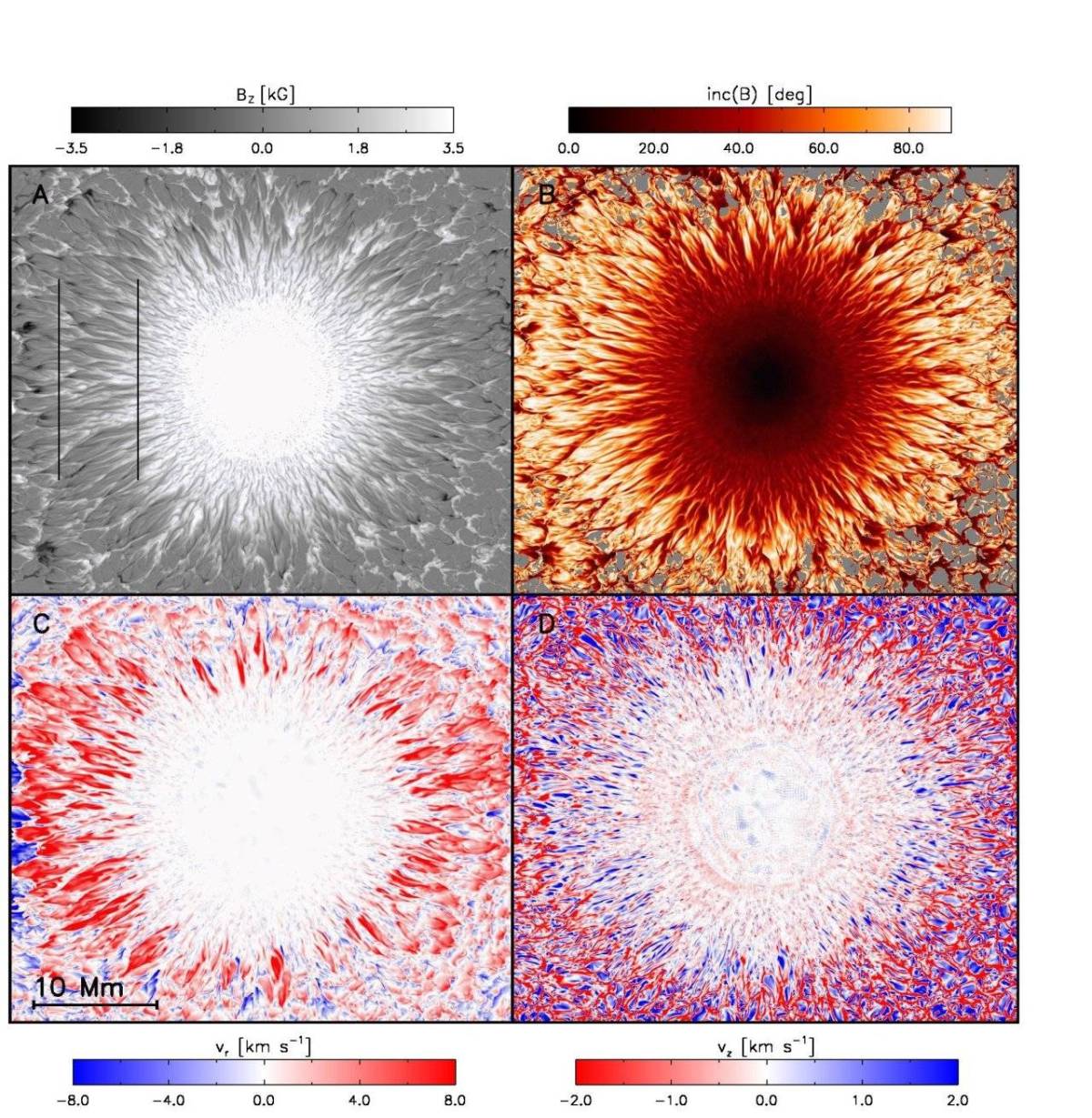}}
        \caption{
          Magnetic field and velocity structure at
          the visible surface for the sunspot on the right side of
          Fig.~1. ({\bf A}) Vertical field component (saturated at $\pm
          3.5\,$kG). ({\bf B}) Inclination angle of the magnetic field
          with respect to the vertical direction (grey indicates
          $\vert B\vert<200\,$G). ({\bf C}) Radial outflow velocity
          (saturated at $\pm 8\,$km$\cdot$s$^{-1}$, red indicates 
          outflows). ({\bf D}) Vertical velocity (saturated at 
          $\pm 2\,$km$\cdot$s$^{-1}$, red indicates downflows).  The
          vertical lines in A indicate the positions of the cuts
          shown in Fig. 4.
        }
\end{figure}  
     
\clearpage
     
\begin{figure}[ht]
  \resizebox{\hsize}{!}{\includegraphics{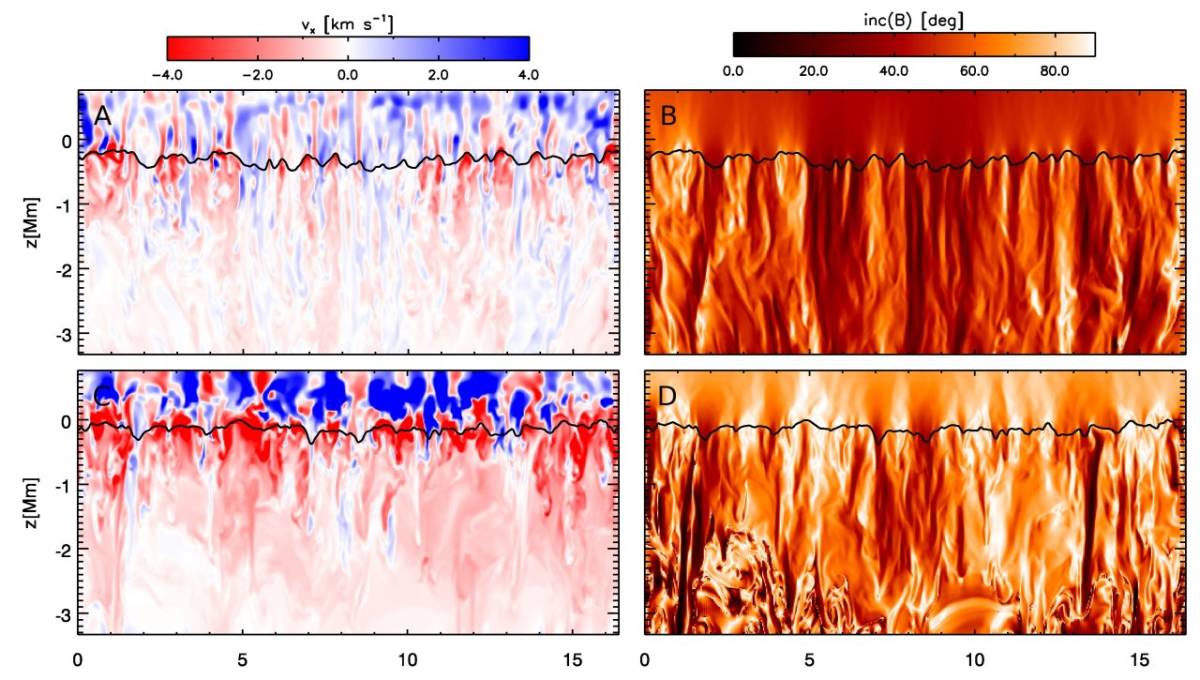}}
  \caption{
    Vertical cuts through the penumbra (indicated by
    the black lines in Fig. 3A). The vertical direction is stretched 
    by a factor of 2. Shown are the
    horizontal velocity component $v_x$ (left) and the field
    inclination (right). ({\bf A, B}) Inner penumbra (right line in
    Fig. 3A). ({\bf C, D}) Outer penumbra (left line in Fig. 3A). The
    color representation of $v_x$ is saturated at $\pm
    4\,$km$\cdot$s$^{-1}$. Black lines indicate the visible surface
    (optical depth unity).
  }
\end{figure}  

\clearpage

\renewcommand{\tot}[1]{\int {#1}\,{\rm d}x{\rm d}y}
\renewcommand{\avr}[1]{\langle{#1}\rangle}

\noindent
{\bf \Large Appendix: Supporting online material (SOM)}
\section{Simulation setup}
The simulation presented here has been carried out with the {\it MURaM}
MHD code \cite{Voegler:2003, Voegler:etal:2005}, with modifications
described in \cite{Rempel:etal:2009a}.  The physics, numerics and
boundary conditions are similar to earlier runs described there, the
primary difference here is the far larger domain size and the initial
magnetic field configuration.

It is currently still out of reach to run an ab-initio simulation of the
formation of an active region, primarily due to the large vertical
extent of the simulation domain required for this purpose. In this study
we focus on a $6.1$ Mm deep domain to study the near-surface structure
of a pair of opposite polarity sunspots. We start with a $98\times
49\times 6.1$ Mm domain of thermally relaxed convection, in which the
$\tau=1$ surface is located about $700$ km beneath the top boundary. The
magnetic field is initialized as a pair of axisymmetric
opposite-polarity sunspots, which are based on the self-similar field
configuration used by \cite{Schuessler:Rempel:2005}. Each spot comprises 
about $1.6\cdot 10^{22}$ Mx magnetic flux, but their initial field strength is
different ($5$ kG and $7$ kG, respectively, at the bottom of the
domain, dropping to about $3$ kG in the near photospheric layers). In the 
course of the simulation, this leads to a pair of
sunspots with a photospheric field strength of about $3.2$ kG (spot on
the left) and $4.2$ kG (spot on the right).  The separation of the two
spots in the middle of the domain is $46$ Mm, about $3$ Mm less than
half of the horizontal extent in the $x$-direction.  Owing to the
periodic horizontal boundary condition this setup allows to study a
variety of different combinations of field strength and inclination
angles: In the $x$-direction we have opposite polarity spots with $3.2$
and $4.2$ kG strength and separations of $46$ and $52$ Mm in between, in
the $y$-direction the magnetic field is less inclined since periodicity
imposes same polarity spots in a distance of $49$ Mm. As a consequence we
can study several realizations of penumbra in one simulation run and
evaluate the robustness of our results.

We ran the simulation for the first hour of simulated time with a rather
low numerical grid resolution of $96\times 96\times 32$ km to get past
initial transients. The second hour was performed at a medium resolution
of $48\times48\times 24$ km and then followed by another $1.6$ hours
with a resolution of $32\times32\times 16$ km (corresponding to
$3072\times 1536\times 384$ grid cells). The results presented here are
based on snapshots near the end of the high-resolution run and partly on
temporal averages over the last hour. The total duration of the run of
about $3.6$ hours is still very short compared to the typical lifetime
of a sunspot, so that the umbral regions are not yet completely thermally
relaxed .  However, the dynamical time scales for the penumbral regions
is much shorter and no significant change or trend of the properties
discussed here has been observed over the $1.6$ hours duration of the high
resolution run.
   
\section{Scales and anisotropy of energy and mass flux}

\subsection{Energy flux}
The vertical energy flux is given by
\begin{equation}
  F_z(x,y,z)=\varrho v_z\left(\varepsilon
             +\frac{p}{\varrho}+\frac{1}{2} v^2\right)+F^M_z
\end{equation}
Here $\varepsilon$ denotes the specific internal energy and $F_z^M$ the
vertical component of the Poynting flux. In the following discussion we
ignore the Poynting flux, which does not exceed a few percent of the total
energy flux. 

To quantify the scale dependence as well as possible anisotropies we
consider here a dimensionless measure for the decorrelation of the
energy flux when mass flux and specific enthalpy are smoothed independently
over a certain length scale $L$. We define here dimensionless
functions of the height $z$ and smoothing length $L$ that quantify the
decorrelation of the energy flux for smoothing in the $x$-direction,
$P_x$, and in $y$-direction, $P_y$:
\begin{eqnarray}
  P_x(z,L) &=& \frac{\tot{\avr{F_z}_L^{x}}}{\tot{F_z}} \nonumber\\
  P_y(z,L) &=& \frac{\tot{\avr{F_z}_L^{y}}}{\tot{F_z}}\;. \label{def:eflux}
  \label{eq:aniso-eflx}  
\end{eqnarray}
Here the quantity $\avr{F_z}_L^{x}$ (and equivalent $\avr{F_z}_L^{y}$) is 
given by
\begin{equation}
  \avr{F_z}_L^x(x,y,z)=\avr{\varrho v_z}_L^x\avr{\varepsilon+\frac{p}{\varrho}
    +\frac{1}{2}v^2}_L^x\;.
\end{equation}
The brackets indicate smoothed quantities (to remove the contributions of
scales smaller than $L$), which are defined through a convolution using a
Gaussian $G_L$ with the length scale $L$ as full width at half maximum. For
example, a 1-dimensional smoothing in the $x$-direction is defined as
\begin{equation}
  \avr{f}_L^x(x,y,z)=\int f(x^{\prime},y,z)\cdot G_L(x-x^{\prime})\,
      {\rm d}x^{\prime} \;.
\end{equation}

Smoothing vertical mass flux and the specific
enthalpy guaranties a balanced mass flux at each scale $L$, provided
that the original mass flux was balanced within the domain.
In general the vertical mass flux is not necessarily balanced within a
subdomain, owing to the presence of vertical oscillations and, possibly,
of flows on scales larger than the subdomain. A meaningful determination
of the convective energy flux requires that these contributions must be
eliminated. We achieve this by subtracting, at each height level, the
density-weighted vertical mean velocity $\bar{v}_z(z)=\tot{\varrho
v_z}/\tot{\varrho}$ from the vertical velocity component before energy
and mass fluxes are computed. As a consequence, we include only the
contribution from motions that overturn within the subdomain boundaries.

\subsection{Mass flux}
We also apply the same method to analyze the scale dependence and anisotropy
of the mass flux. To this end we smooth $\varrho v_z$ as described above and
compute the quantities $Q_x$ and $Q_y$ in the following way:
\begin{eqnarray}
  Q_x(z,L) &=& \frac{\tot{{\vert\avr{\varrho v_z}_L^{x}}\vert}}
    {\tot{\vert \varrho v_z \vert}} \nonumber \\ 
  Q_y(z,L) &=& \frac{\tot{{\vert\avr{\varrho v_z}_L^{y}}\vert}}
    {\tot{\vert \varrho v_z \vert}} \label{def:mflux}
\end{eqnarray}
Since the mass flux is balanced, i.e. $\tot{\varrho v_z}=0$, we consider here
the absolute value. Our analysis will therefore provide the typical scale
on which most of the mass flux turns around.

\subsection{Results}
\begin{figure*}
  \centering 
  \resizebox{\hsize}{!}{\includegraphics{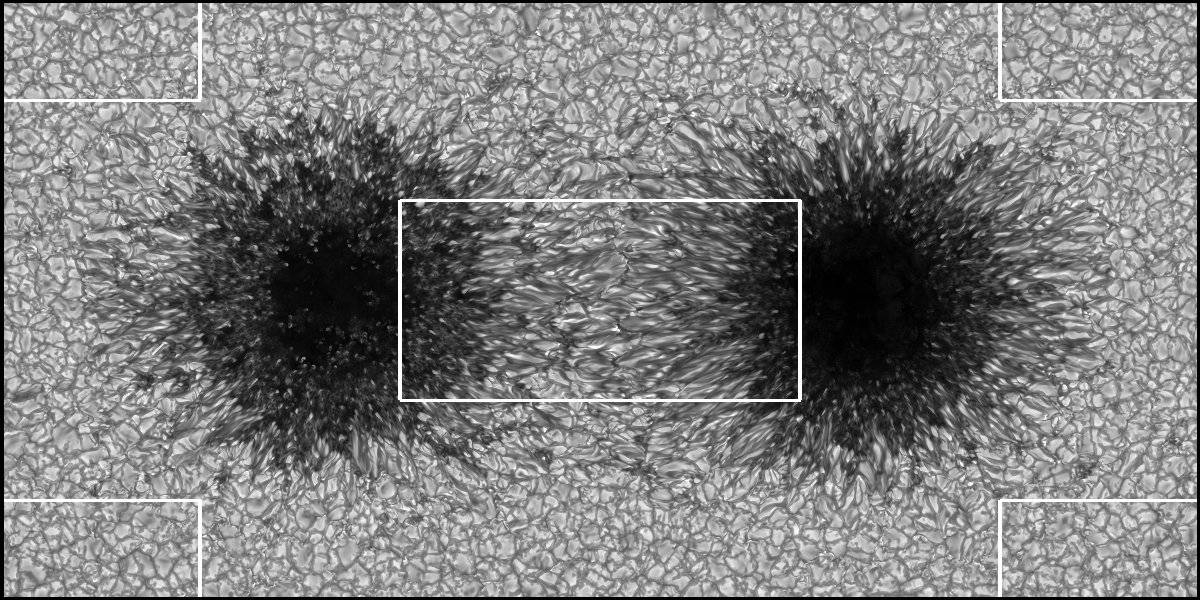}}
  \caption{White boxes indicate the domains for which we compare the 
    properties of energy and mass flux. The central box encloses most of the
    penumbral region between both spots, the box near the corners encloses
    a region of equal size with almost undisturbed granulation.}
  \label{f1}
\end{figure*}

\begin{figure*}
  \centering 
  \resizebox{\hsize}{!}{\includegraphics{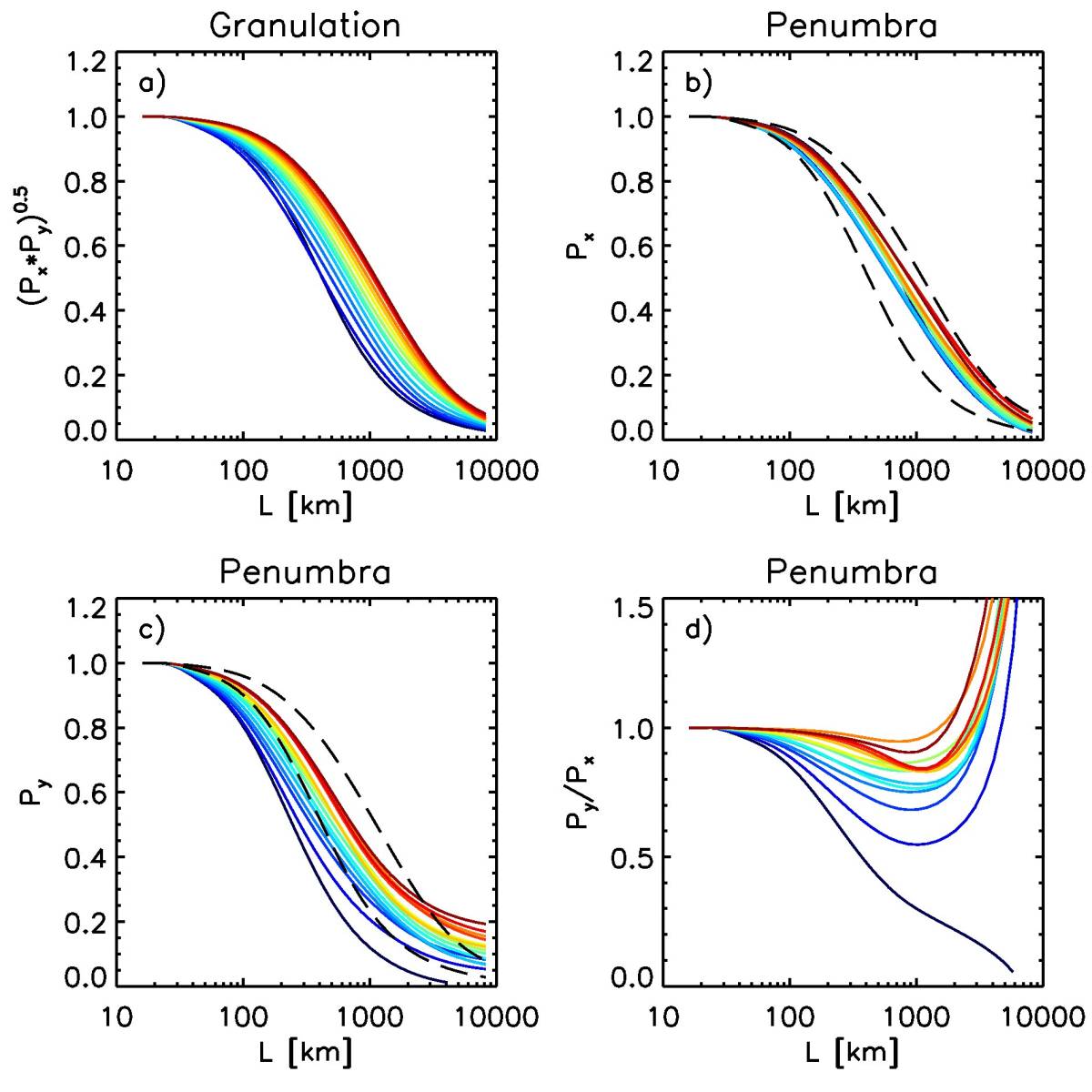}}
  \caption{Scale dependence and anisotropy of energy flux in solar
    granulation and penumbral region. Panel a) shows undisturbed
    granulation for reference.  Different height levels are color-coded
    ranging from the photosphere (dark blue) to the bottom of the domain
    (dark red), spanning a total of $4.8$ Mm. The black dashed lines in
    panels b) and c) correspond to the dark blue (near photosphere) and
    dark red (near bottom of domain) curves in panel a). Panels b) and
    c) display the corresponding quantities for the penumbral region, panel
    d) the derived anisotropy of the energy flux with respect to the $x$
    and $y$ direction.  }
  \label{f2}
\end{figure*}

To compare the properties of the energy flux in a penumbral region and
almost undisturbed granulation we apply this procedure to two subdomains
indicated in Fig.~\ref{f1}.  Both regions have about $32\times 16$ Mm
horizontal extent (note that both boxes are contiguous owing to the
periodic boundary condition). Fig.~\ref{f2} presents the scale dependence
and anisotropy of the energy flux comparing the granulation and
penumbral region. Panel a) shows the geometric average of $P_x$ and
$P_y$ for granulation. Different height levels are color-coded, with
depth increasing from blue toward red. The uppermost height level
corresponds to the position at which the average pressure scale height
is $200$ km, the distance between the height levels is $320$ km. The
total range covered in the vertical direction is about $4.8$ Mm,
excluding the lower most $500$ km that are affected by the boundary
condition. With increasing smoothing length scale $L$ (thus eliminating
the contribution from small scales) the remaining convective energy flux
decreases and drops by a factor of about $2$ when $L$ reaches a value of
about $400$ km in the near photospheric layers. For the deeper layers
the corresponding curves are shifted toward larger values of $L$
(approximately $L\sim H_p$). The anisotropy, defined as $P_y/P_x$, 
(not shown here) is very
close to $1$ with fluctuations of up $20\%$ on larger scales. Panels b)
and c) present the results from applying the same procedure to the
penumbral region between the two spots as indicated by the central
white box in Fig.~\ref{f1}. While the overall shape of the curves
remains the same, some distinct differences occur for $P_x$ and
$P_y$. In the case of $P_x$, the depth dependence is strongly reduced and
all curves almost coincide with a curve corresponding to that from 
about $2-3$ Mm
depth in the case of granulation. On the other hand the depth dependence 
of $P_y$ is similar to that of granulation, but overall the scales are 
reduced by
a factor of about $2$. Panel d) shows the anisotropy defined through the
ratio $P_y/P_x$. For scales of less than about $2000$ km, the deep
layers remain close to isotropic, while the anisotropy reaches values of
about $0.25$ in the near photospheric layers. According to our
definition of the anisotropy values $<1$ indicate smaller scales in the
$y$ as compared to the $x$-direction (i.e.  the quantity
$\tot{\avr{F_z}_L^{y}}$ falls off quicker than $\tot{\avr{F_z}_L^{x}}$
with increasing $L$).  For scales larger than $2000$ km the deeper layers
have anisotropies $>1$, indicating a contribution from large scale flows
more coherent in the $y$ direction. In the deep layers contributions
from large scales in $P_y$ remain at about $20\%$ compared to about
$10\%$ in undisturbed granulation. The difference of $10\%$ can be
attributed to the presence of the large scale Evershed/moat flow system
in the penumbra region.

\begin{figure*}
  \centering 
  \resizebox{\hsize}{!}{\includegraphics{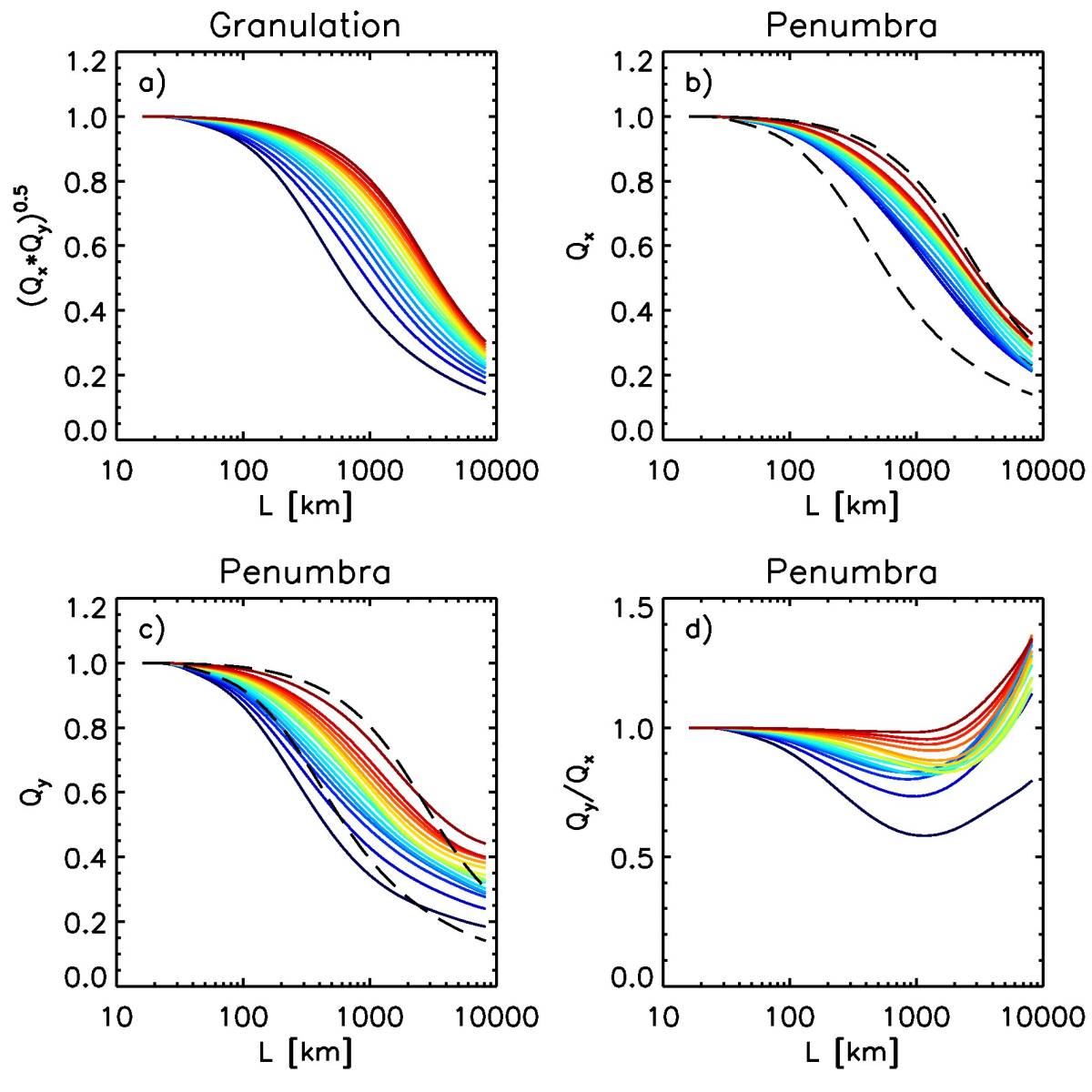}}
  \caption{Same as Fig.~\ref{f3} for the mass flux as defined by Eq. 
    (\ref{def:mflux}). }
  \label{f3}
\end{figure*}

On a qualitative level the behavior of the  mass flux (Fig.~\ref{f3}) is not
distinctly different from that of the energy flux. 
Overall the typical scale for the overturning of mass
is larger than the typical scale for the energy transport in both
granulation and penumbra. The degree of anisotropy of the mass flux is
reduced compared to the anisotropy of the energy flux. The increase
of anisotropy on larger scales is more pronounced on all height
levels. In the deep layers the Evershed/moat flow contribution to the
mass flux is about $10\%$ to $15\%$ (values of $Q_y$ remain around
$45\%$ as compared to $30\%$ in the case of granulation).

While the penumbral region shows distinct differences from granulation
in terms of aniso\-tropy in height levels extending several Mm downward,
we find no significant change in the relative contribution of large and
small scales in the overall energy transport. The presence of large
scale contributions is more prominent in the mass flux, but not
substantially enlarged compared to granulation.

\begin{figure*}
  \centering 
  \resizebox{\hsize}{!}{\includegraphics{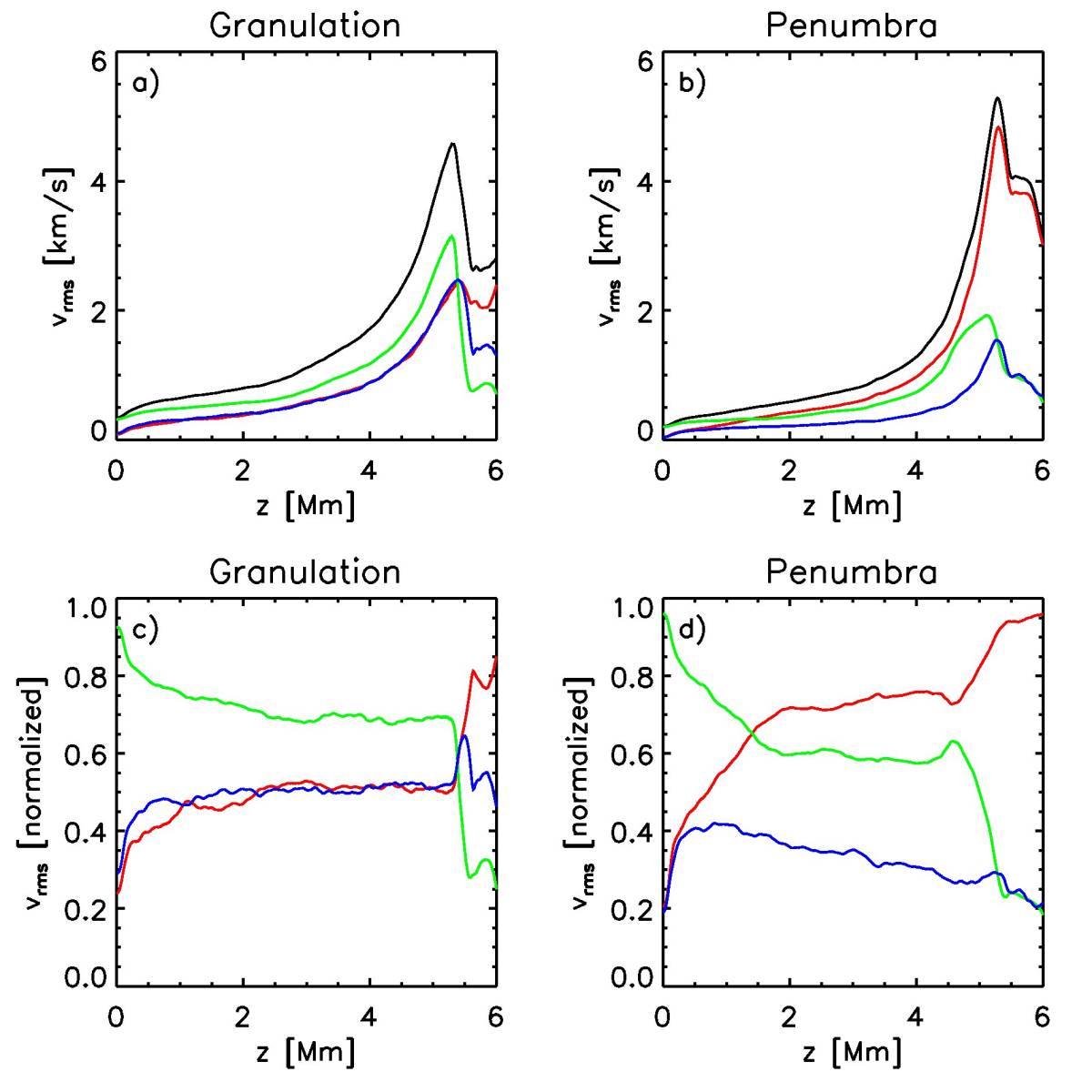}}
  \caption{Comparison of rms velocities in granulation and
    penumbra. Panels a) and b) show $v_{\rm rms}$ (black) $v_{x\,\rm
    rms}$ (red), $v_{y\,\rm rms}$ (blue) and $v_{z\,\rm rms}$ (green)
    for granulation and penumbra, respectively. Panels c) and d) present
    $v_{x\,\rm rms}$, $v_{y\,\rm rms}$ and $v_{z\,\rm rms}$ normalized
    by $v_{\rm rms}$.}
  \label{f4}
\end{figure*}

The similarities between energy and mass transport in granulation and
penumbra are also evident from the height dependence of rms
velocities (Fig.~\ref{f4}). The computation of the rms
velocities in the penumbra is based on a smaller subdomain of half the
size as indicated in Fig.~\ref{f1} to exclude umbral regions which
would lower the overall rms velocity (for computing the rms velocities a
balanced mass flux within the subdomain is less important). We present
in panels a) and b) $v_{\rm rms}$ (black) $v_{x\,\rm rms}$ (red), 
$v_{y\,\rm rms}$ (blue) and $v_{z\,\rm rms}$ (green) for granulation and 
penumbra, respectively; panels c) and d) display the rms velocities 
normalized by the total rms. A comparison for the profile of $v_{\rm rms}$ 
between granulation and penumbra does not reveal a significant difference,
except for slightly larger velocities in the photosphere and a steeper
increase toward the photosphere. In the case of
granulation vertical motions contribute about $50\%$ to the kinetic
energy, while both horizontal components contributing about
$25\%$. These contributions are almost independent of depth except near
the bottom where $v_z$ is preferred over horizontal motions due to the
boundary condition. In the case of the penumbra the kinetic energy is
dominated by the component in the direction of filaments ($x$) which
contributes about $55\%$, the contribution of vertical motions is
reduced to about $35\%$ while horizontal motions perpendicular to
filaments contribute only about $10\%$ in the near photospheric layers.

Overall we do not see a strong indication for a fundamental difference
between energy and mass transport in granulation and penumbra, except
for the anisotropy introduced by the presence of strong horizontal
magnetic field.  Scales in the direction of the horizontal field become
less dependent on the pressure scale height, while scales perpendicular
to the preferred field direction remain strongly dependent on pressure
scale height and are reduced. As a consequence energy and mass transport
become more anisotropic toward the surface. The anisotropy leads to a
reduction of horizontal motions perpendicular to the field direction,
while the component parallel to the field is increased. The total rms
velocity does not show a significant change, indicating that the kinetic
energy of convective motions remains similar, it is just differently
distributed among the $x$, $y$ and $z$ direction.  These similarities
strongly indicate that driving the Evershed flow does not require
physical processes that go beyond the combination of convection and
anisotropy introduced by the magnetic field.

\clearpage

\end{document}